# Application of the Finite Field Coupled Cluster Method to Calculate Molecular Properties Relevant to Electron Electric Dipole Moment Searches

M Abe[1], V S Prasannaa[2,3], B P Das[4]

[1] Tokyo Metropolitan University, 1-1, Minami-Osawa, Hachioji City, Tokyo 192-0397, Japan
[2] Indian Institute of Astrophysics, Koramangala 2nd block, Bangalore- 560034, India
[3] Department of Physics, University of Calicut, Malappuram, Kerala- 673 635, India
[4] Department of Physics, School of Science, Tokyo Institute of Technology, 2-12-1-H86, Ookayama, Meguro-Ku, Tokyo-153-8550, Japan

**Abstract:**
Heavy polar diatomic molecules are currently among the most promising probes of fundamental physics. Constraining the electric dipole moment of the electron (eEDM), in order to explore physics beyond the Standard Model, requires a synergy of molecular experiment and theory. Recent advances in experiment in this field have motivated us to implement a finite field coupled cluster approach (FFCC). This work has distinct advantages over the theoretical methods that we had used earlier in the analysis of eEDM searches. We used relativistic FFCC to calculate molecular properties of interest to eEDM experiments, that is, the effective electric field ($E_{eff}$) and the permanent electric dipole moment (PDM). We theoretically determine these quantities for the alkaline earth monofluorides (AEMs), the mercury monohalides (HgX), and PbF. The latter two systems, as well as BaF from the AEMs, are of interest to eEDM searches. We also report the calculation of the properties using a relativistic coupled cluster approach with single, double, and partial triples' excitations, which is considered to be the gold standard of electronic structure calculations. We also present a detailed error estimate, including errors that stem from our choice of basis sets, and higher order correlation effects.

**Introduction:**

In recent years, heavy polar diatomic molecules have become extremely important probes of fundamental physics [1,2]. Using this non-accelerator approach, one constrains exotic properties like electric dipole moments of leptons, and the nuclear anapole moment. Recent experiments, in combination with relativistic molecular calculations, to set an upper bound on the parity (P) and time reversal (T) violating electric dipole moment of the electron (eEDM) have provided us with constraints for physics beyond the Standard Model at the TeV and higher energy scales [3]. This upper bound can also help shed light on the baryon asymmetry in our universe [4,5]. The current best limit on eEDM is set by ThO [6], followed by HfF$^+$[7] and YbF [8]. New experiments using molecules like BaF are underway [9], in a bid to set better limits. These advances are made possible not only due to developments on the experimental front, but also in theory. This is because eEDM is obtained by combining the experimentally observed shift in energy of a molecule in a particular state, due to eEDM, and the effective electric field, $E_{eff}$. The latter is a relativistic effect [10], and can only be calculated using a relativistic many-body theory. The relativistic coupled cluster method (RCC method), which is equivalent to an all-order relativistic many body perturbation theory for a given level of particle-hole excitation, plays a pivotal role in such situations. This is because high precision calculations are required, which in combination with experiments that are highly sensitive, to explore physics beyond the Standard Model.

We had developed a RCC method, and applied it to calculate $E_{eff}$ and the PDM of YbF [11]. Later, we used the same method to identify HgX as promising candidates for future eEDM experiments [12], because they had extremely large values of $E_{eff}$. This directly translates into a significant improvement in the sensitivity of an eEDM experiment.

The same theoretical technique was also adapted for determining the PDMs of the AEMs [13,14], which are important because most of them have the potential to be laser cooled, and hence can be used in several high precision experiments. In fact, SrF was the first molecule to be laser-cooled [15], and it might be used for a test of parity violation

[16]. CaF has also been laser cooled [17], and high precision experiments may be performed on it, subsequently [18]. BaF is not only suitable for eEDM experiment [9], but has also been reported as a promising candidate to probe the nuclear anapole moment [19-21]. Subsequently, we also computed the PDMs of HgX [22], which plays a role in the sensitivity of an eEDM experiment, via the polarizing field. Detailed discussions on the importance of these systems can be found in our earlier works [12,13,14,22].

Our RCC method was based on the Dirac-Coulomb Hamiltonian and yielded accurate results, but we had made an approximation in the calculation of the expectation values in using the coupled cluster singles and doubles (CCSD) wave function, since its expansion does not terminate. We only considered leading contributions of the cluster operators (T), which are the linear terms, for the expectation values (LECC), while the cluster amplitudes were optimized, employing the full CCSD method [11].

In this paper, we go beyond the LECC approximation, by adapting the finite field perturbation theory to the relativistic CCSD method (FFCC), and applying it to the effective electric fields and the PDMs of AEMs, HgX, and to another eEDM candidate, PbF. Since the expression for the correlation energy terminates (due to the Hamiltonian having only one and two body terms), the energy derivative in FFCC approach requires no approximation to be made for a given level of particle-hole excitation, unlike the LECC approximation. Later in this work, we also apply the finite field version of the relativistic CCSD(T) method [23], the gold standard of electronic structure calculations [24], which also includes partial triples to $5^{th}$ order, besides single and double excitations.

An accurate method to calculate these properties becomes useful, since this approach also aids in indirectly looking for the importance of the correlation effects arising from the non-linear terms (in T), in their expectation value expression. In some molecules, where the correlation effects are large (PbF's PDM changes by 20 percent when correlations are taken into account in LECC approximation, while HgI's reduces by half), this becomes all the more important.

**Theory and Methodology:**

In our FFCC approach, we carefully select the perturbation parameters and number of grid points to suppress the errors from numerical derivatives. We also estimate complete basis sets (CBS) limits for the PDM and $E_{eff}$ values of AEMs at the CCSD level using double zeta (DZ), triple zeta (TZ), and quadruple zeta (QZ) basis sets. The accuracy of the LECC approximation is tested comparing with the FFCC results. The deviations from FFCC method are reasonably small in most cases, in the molecules presently calculated. Hence, we demonstrate that the LECC method is a computationally fast and accurate tool to estimate the values of $E_{eff}$ and PDM for some simple systems, such as doublet sigma electronic states.

In the LECC approximation [11,12,13,14,22, and references therein], we calculated the expectation values of a one-body operator $\hat{O}$ at the CCSD level as follows:

$$\frac{\langle \Psi|\hat{O}|\Psi \rangle}{\langle \Psi|\Psi \rangle} = \langle \Phi_0|e^{\hat{T}^\dagger}\hat{O}_N e^{\hat{T}}|\Phi_0 \rangle_C + \langle \Phi_0|\hat{O}|\Phi_0 \rangle \quad (1)$$

$$\approx \langle \Phi_0|\left(1+\hat{T}_1+\hat{T}_2\right)^\dagger \hat{O}_N \left(1+\hat{T}_1+\hat{T}_2\right)|\Phi_0 \rangle_C + \langle \Phi_0|\hat{O}|\Phi_0 \rangle \quad (2)$$

Here, $\Phi_0$ and $\Psi$ denotes the unperturbed Dirac-Fock (DF) and coupled cluster wave functions. $\hat{T}$ is the cluster operator and $\hat{O}_N$ is the normal ordered version of $\hat{O}$, defined by $\hat{O}-\langle \Phi_0|\hat{O}|\Phi_0 \rangle$. The subscript C in Eq.s (1) and (2) indicates that only the connected terms are taken into account [25]. As we go from Eq. (1) to (2), we neglect the excitations higher than doubles ($\hat{T}_3, \hat{T}_4,...$), and the nonlinear terms arising from the expansions of Eq. (1) in T$_1$ and T$_2$ (such as $\hat{T}_1^2, \hat{T}_1\hat{T}_2$).

The FFCC method is an alternate approach to calculate properties, where we calculate energy derivatives of a perturbation parameter, $\lambda$, numerically, instead of directly calculating expectation values. We consider a perturbation, $\hat{O}$, which is a one-body operator, to the unperturbed Hamiltonian, H:

$$\hat{H}(\lambda) = \hat{H} + \lambda\hat{O} \quad (3)$$

and define its normal ordered version, $\hat{H}_N(\lambda)$ as follows.

$$\hat{H}_N(\lambda) = \hat{H}_N + \lambda \hat{O}_N. \quad (4)$$

Here, we choose the Dirac-Coulomb Hamiltonian for $\hat{H}$. The correlation energy, $E_{corr}(\lambda)$, using the linked cluster theorem [25], is given by:

$$E_{corr}(\lambda) = \langle \Phi_0 | \left( \hat{H}_N(\lambda) e^{\hat{T}(\lambda)} \right)_C | \Phi_0 \rangle \quad (5)$$

In this work, we use the one-particle orbitals obtained by the *unperturbed* DF calculations and hence, the bra and ket in Eq. (5) is written using the unperturbed DF wave function ($\Phi_0$) instead of the perturbed one ($\Phi_0(\lambda)$). We need to adopt this strategy because the DF program in UTChem is constructed under the restriction of Kramers' symmetry. Therefore, we cannot add T violating perturbations, such as eEDM, at the DF level. However, the effect of orbital relaxation would be recovered at the perturbed CCSD level, as shown in [26]. The computational cost of DF and integral transformation is also drastically reduced under the Kramers' restriction, and hence, we consider the perturbation at the CCSD level by adding the integrals associated with the perturbation to the original one-electron operator integrals.

We obtain the energy derivative with respect to the parameter $\lambda$, using the two-point and six-point central difference methods, which are given by:

$$\left. \frac{dE_{corr}(\lambda)}{d\lambda} \right|_{\lambda=0} \approx \frac{E_{corr}(\lambda) - E_{corr}(-\lambda)}{2\lambda} \quad (6)$$

$$\left. \frac{dE_{corr}(\lambda)}{d\lambda} \right|_{\lambda=0} \approx 3 \frac{E_{corr}(\lambda/4) - E_{corr}(-\lambda/4)}{\lambda} + \frac{E_{corr}(3\lambda/4) - E_{corr}(-3\lambda/4)}{15\lambda} - 3 \frac{E_{corr}(\lambda/2) - E_{corr}(-\lambda/2)}{5\lambda} \quad (7)$$

respectively. The equations given above only represent the electron correlation effects. The DF contribution should be added to obtain the total contribution to a property, as shown below:

$$\frac{\langle \Psi | \hat{O} | \Psi \rangle}{\langle \Psi | \Psi \rangle} = \left. \frac{dE_{corr}(\lambda)}{d\lambda} \right|_{\lambda=0} + \langle \Phi_0 | \hat{O} | \Phi_0 \rangle. \quad (8)$$

In this work, we calculate $E_{\text{eff}}$ by setting:

$$\hat{O} \equiv \hat{H}_{\text{eEDM}}/d_e = -2ic\sum_j^{N_e} \beta\gamma_5 p_j^2 \quad (9)$$

and

$$\lambda \equiv d_e. \quad (10)$$

Here $\beta$ and $\gamma_5$ are the one of the Dirac matrices and $p$ is the momentum operator, while $d_e$ is the absolute value of eEDM.

Similarly, we obtained the PDM by setting:

$$\hat{O} \equiv -\sum_j^{N_e} z_j \quad (11)$$

and

$$\lambda \equiv F_z. \quad (12)$$

We fix the molecular axis to the z-axis, and $F_z$ is the perturbation parameter, indicating the strength of the external electric field along the molecular axis. We also add the nuclear contribution to the electronic part, in order to obtain the final value for PDM.

Within the FFCC method, we need to calculate CCSD energies several times (for each $\hat{O}$, and various values of $l$ independently), thereby increasing the computational cost. In contrast, the LECC approximation requires only a single CCSD calculation, with the unperturbed Hamiltonian, after which we calculate expectation values of any one-body operator using the information of the unperturbed cluster amplitudes.

We perform the DF calculations and obtain the one-body integrals for the properties, using the modified UTChem code [27,28]. We calculate the correlation energies for various values of $\lambda$ ($10^{-3}$, $10^{-4}$, $10^{-5}$, and $10^{-6}$) at the CCSD level using the Dirac08 code [29]. We employ the same uncontracted Gaussian-type basis sets, as in our previous works, to compare the current results with those obtained by using the LECC approximation. The details of the basis sets are: cc-pVnZ for the halides, Be, Mg and Ca, Dyall's cnV basis sets for Hg and Pb, and a combination of Dyall and Sapporo basis sets to include diffuse functions for Sr and Ba; n is the cardinal number, and is 2 for DZ, 3 for

TZ, and 4 for QZ basis sets (for details about the basis sets, refer [12,13,14], and references therein). We use the following bond lengths (in Angstrom): BeF: 1.361, MgF: 1.75, CaF: 1.967, SrF: 2.075 [30,31], BaF: 2.16 [32,33], PbF: 2.06 [34], HgF: 2.00686, HgCl: 2.42, HgBr: 2.62, and HgI: 2.81 [35,36]. In our coupled cluster calculations, we do not freeze any of the occupied orbitals. We also do not impose any cut-off on the virtuals.

**Results and Discussions:**

TABLE I: PDMs (Debye) and effective electric fields (GV/cm) of AEMs at DF, LECC, and FFCC levels. The percentage difference between LECC and FFCC methods are shown in 'Diff'.

|  | PDM (Debye) | | | | $E_{\text{eff}}$ (GV/cm) | | | |
|---|---|---|---|---|---|---|---|---|
|  | DF | LECC | FFCC | Diff (%) | DF | LECC | FFCC | Diff (%) |
| BeF (DZ) | 1.32 | 0.93 | 1.01 | 7.9 | 0.002 | 0.003 | 0.003 | 0.0 |
| BeF (TZ) | 1.31 | 1.06 | 1.12 | 5.4 | 0.002 | 0.004 | 0.004 | 0.0 |
| BeF (QZ) | 1.30 | 1.10 | 1.15 | 4.3 | 0.003 | 0.005 | 0.005 | 0.0 |
| MgF (DZ) | 3.21 | 2.84 | 2.91 | 2.4 | 0.04 | 0.06 | 0.06 | 0.0 |
| MgF (TZ) | 3.21 | 3.02 | 3.08 | 1.9 | 0.05 | 0.06 | 0.06 | 0.0 |
| MgF (QZ) | 3.16 | 3.07 | 3.13 | 1.9 | 0.05 | 0.07 | 0.07 | 0.0 |
| CaF (DZ) | 2.89 | 3.01 | 3.07 | 2.0 | 0.16 | 0.23 | 0.23 | 0.0 |
| CaF (TZ) | 2.82 | 3.13 | 3.17 | 1.3 | 0.18 | 0.27 | 0.27 | 0.0 |
| CaF (QZ) | 2.77 | 3.16 | 3.19 | 0.9 | 0.19 | 0.28 | 0.28 | 0.0 |
| SrF (DZ) | 2.83 | 2.95 | 3.02 | 2.3 | 1.33 | 1.91 | 1.99 | 4.0 |
| SrF (TZ) | 2.95 | 3.42 | 3.46 | 1.2 | 1.51 | 2.14 | 2.12 | -0.9 |
| SrF (QZ) | 3.01 | 3.60 | 3.62 | 0.6 | 1.54 | 2.17 | 2.16 | -0.5 |
| BaF (DZ) | 2.42 | 2.69 | 2.77 | 2.9 | 4.58 | 6.48 | 6.42 | -0.9 |
| BaF (TZ) | 2.28 | 3.00 | 2.96 | -1.4 | 4.83 | 6.65 | 6.60 | -0.8 |
| BaF (QZ) | 2.65 | 3.40 | 3.41 | 0.3 | 4.80 | 6.50 | 6.46 | -0.6 |

TABLE II: PDMs (Debye) and effective electric fields (GV/cm) of HgX and PbF at the DZ level of basis.

|  | PDM (Debye) | | | | $E_{\text{eff}}$ (GV/cm) | | | |
|---|---|---|---|---|---|---|---|---|
|  | DF | LECC | FFCC | Diff (%) | DF | LECC | FFCC | Diff (%) |
| HgF | 3.96 | 2.61 | 2.92 | 10.6 | 104.25 | 115.42 | 116.37 | 0.8 |
| HgCl | 4.23 | 2.72 | 2.96 | 8.1 | 103.57 | 113.56 | 114.31 | 0.7 |
| HgBr | 4.40 | 2.36 | 2.71 | 12.9 | 97.89 | 109.29 | 109.56 | 0.2 |
| HgI | 3.91 | 1.64 | 2.06 | 20.4 | 96.85 | 109.30 | 109.56 | 0.2 |
| PbF | 4.42 | 3.72 | 3.88 | 4.1 | 40.20 | 37.24 | 37.91 | 1.8 |

Tables I and II show the values of PDM and the absolute values of $E_{\text{eff}}$ for various AEMs and some halides, respectively. In all the cases, we have rounded off the values to two decimal places except for $E_{\text{eff}}$ of BeF, whose values are less than $10^{-2}$. This rounding off is done, because accuracy of quantum chemical calculations for these properties are usually not higher than this level. For all the data pertaining to the FFCC approach in Table I, the results of two and six-point formulae agree well beyond three decimal places. Hence, for the heavier systems in Table II, we obtained the values using only the two-point formula, to avoid expensive computations. We also tested $\lambda$-dependence between $10^{-3}$ and $10^{-6}$ in FFCC calculations, in steps of $10^{-1}$. The two properties do not change with change in $\lambda$, at least up to three decimal places. Hence within the selected range of $\lambda$, the obtained values of PDM and $E_{\text{eff}}$ converge. As we go from DZ through QZ basis sets in Table I, we see that the PDMs converge for all AEMs, except BaF. For $E_{\text{eff}}$, CaF and SrF converge clearly, BeF increases consistently, while MgF and BaF do not converge.

Next, we shall look at the difference between the LECC and FFCC methods. For all the molecules, the LECC values of $E_{\text{eff}}$ do not deviate significantly from their FFCC counterparts. The larger deviations are found in poorer basis sets (like DZ). In contrast, PDM displays a different trend, with the largest difference is 20% for HgI (DZ). This indicates that correlation effects (both linear and non-linear in T) in the PDMs of HgX are very pronounced. However, this deviation may possibly decrease if we use a higher basis set, as the deviations using the QZ basis sets are less than those with DZ in AEMs. In most cases, 'Diff' is positive (except BaF(TZ)). This means that the non-linear contributions recovered by FFCC method increase the values of PDM in the molecules considered. For the AEMs, at the QZ level, the importance of the non-linear contribution reduces in heavier systems from BeF (about four percent) to BaF (about 0.3 percent). The PDM does not change significantly even for PbF.

We shall briefly comment on the change in the predicted experimental sensitivity, due to the FFCC approach, by taking HgF as an example. A change of 0.8 percent in $E_{\text{eff}}$, from LECC to FFCC, translates to a change in the statistical uncertainty in an eEDM experiment by the same amount, with all other factors in the figure of merit being the same. This is due to $E_{\text{eff}}$ occurring *outside* the square root, unlike the number of molecules or the integration time, in the expression for statistical uncertainty. A change of 0.8 percent in $E_{\text{eff}}$, combined with about 10 percent in the case of the PDM, results in a percentage fraction difference of about 11, in the so-called sensitivity parameter, which is the ratio of $E_{\text{eff}}$ to the polarizing electric field.

Since the correlation contributions to the two properties are constructed from correlation energies in the FFCC method, we performed a complete basis set extrapolation (CBS) for the correlation parts of these properties. Note that, however, the DF part is not extrapolated, since it is not an energy derivative. We use a polynomial type CBS scheme, which contains two parameters in the function,

$$E(N) = E_{CBS} + AN^{-3}, \quad (12)$$

where $N$ is the cardinal number, i.e. N=2, 3, 4 for DZ, TZ, and QZ basis sets, respectively [37]. To determine $E_{CBS}$ we used the correlation energies obtained at the TZ and QZ-FFCC level as follows,

$$E_{CBS} = \frac{4^3 E_{QZ} - 3^3 E_{TZ}}{4^3 - 3^3}, \quad (13)$$

similar to the relativistic calculation by Gomes et al [38]. Figures 1-10 show the correlation parts of $E_{\text{eff}}$ and PDM ($\Delta E_{\text{eff}}$ and $\Delta PDM$) with respect to N and the extrapolated functions, which provide CBS limit, when N approaches infinity. We also plot the corresponding data obtained by LECC for comparison. In most cases, we observe that the CBS properties do converge, and FFCC results with the QZ basis sets are already close to the CBS values, as shown numerically in Table III. We could also qualitatively confirm that the extrapolated functions are reasonably good, by observing that the DZ FFCC results are close to the function. In our calculation, the agreements of DZ-FFCC data to the functions seem better in PDM, rather than $E_{\text{eff}}$. We also observe that for MgF, up to the QZ level of basis, the effect of correlations is to reduce the PDM. However, at the CBS level, we observe that correlations actually increase it.

TABLE III: The correlation parts of PDMs (Debye) and effective electric fields (GV/cm) of AEMs, at DZ, TZ, QZ, and CBS levels with FFCC method.

| Molecule | Property | DZ | TZ | QZ | CBS |
|---|---|---|---|---|---|
| BeF | $E_{\text{eff}}$ | 0.001 | 0.0015 | 0.0021 | 0.0026 |
|     | PDM     | -0.31 | -0.19 | -0.15 | -0.12 |
| MgF | $E_{eff}$ | 0.015 | 0.018 | 0.021 | 0.023 |
|     | PDM     | -0.30 | -0.13 | -0.03 | 0.05  |
| CaF | $E_{eff}$ | 0.069 | 0.081 | 0.086 | 0.089 |
|     | PDM     | 0.18  | 0.35  | 0.42  | 0.47  |
| SrF | $E_{eff}$ | 0.62  | 0.62  | 0.62  | 0.62  |
|     | PDM     | 0.19  | 0.51  | 0.61  | 0.68  |
| BaF | $E_{eff}$ | 1.84  | 1.77  | 1.66  | 1.57  |
|     | PDM     | 0.35  | 0.68  | 0.76  | 0.81  |

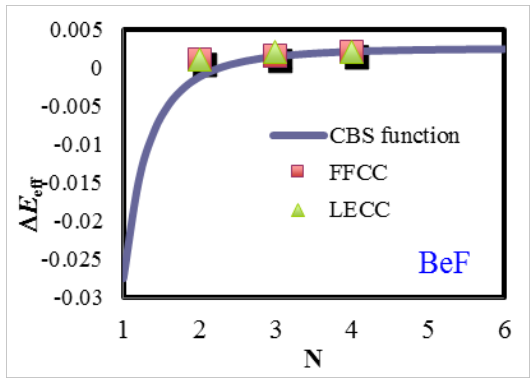

Figure 1 Correlation part of $E_{eff}$ ($\Delta E_{eff}$), in GV/cm, vs cardinal number N (no units) in BeF.

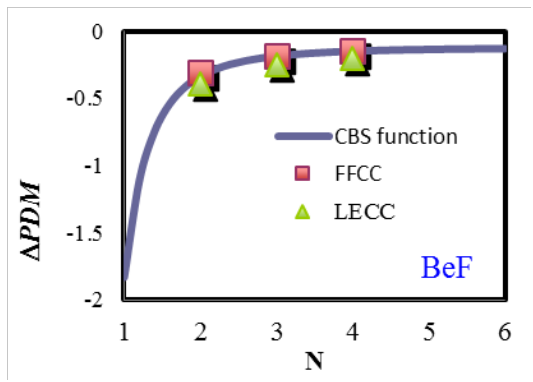

Figure 2. Correlation part of PDM ($\Delta$PDM), in Debye, vs cardinal number N (no units) in BeF.

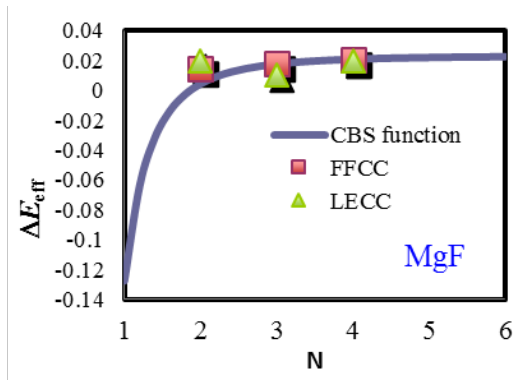

Figure 3 Correlation part of $E_{eff}$ ($\Delta E_{eff}$), in GV/cm, vs cardinal number N (no units) in MgF.

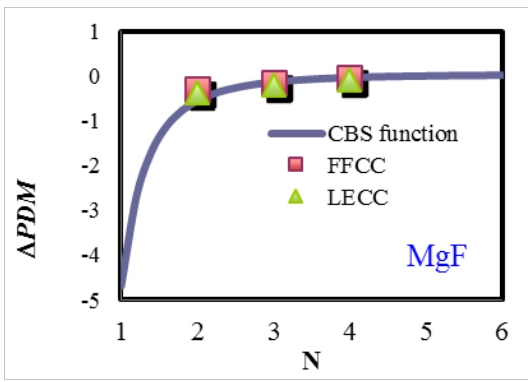

Figure 4. Correlation part of PDM ($\Delta$PDM), in Debye, vs cardinal number N (no units) in MgF.

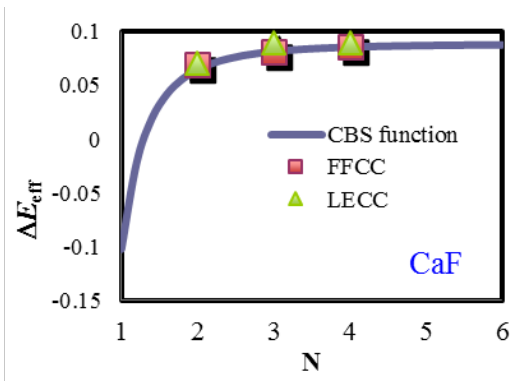

Figure 5 Correlation part of $E_{eff}$ ($\Delta E_{eff}$), in GV/cm, vs cardinal number N (no units) in CaF

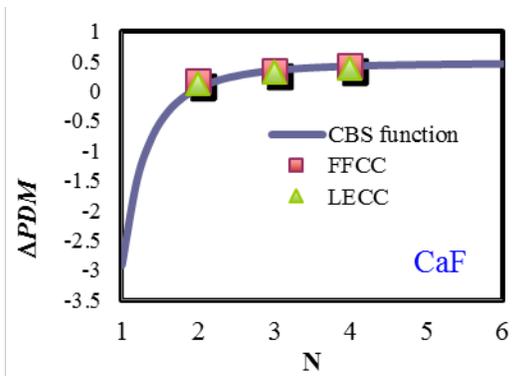

Figure 6. Correlation part of PDM (ΔPDM), in Debye, vs cardinal number N (no units) in CaF.

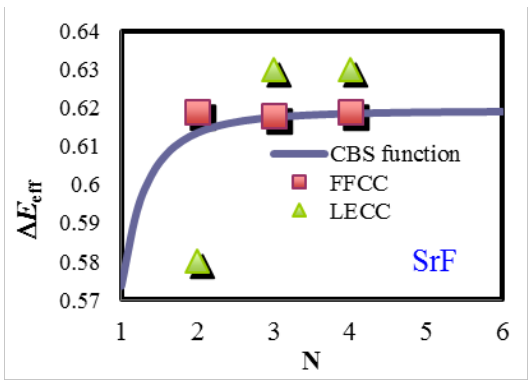

Figure 7. Correlation part of $E_{\text{eff}}$ ($\Delta E_{\text{eff}}$), in GV/cm, vs cardinal number N (no units) in SrF.

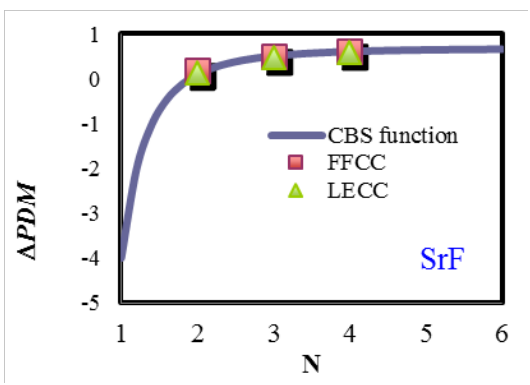

Figure 8. Correlation part of PDM (ΔPDM), in Debye, vs cardinal number N (no units) in SrF

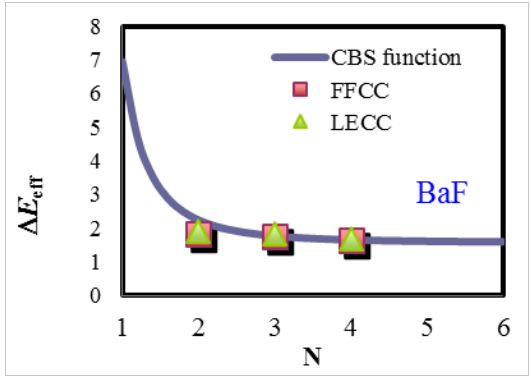

Figure 9. Correlation part of $E_{eff}$ ($\Delta E_{eff}$), in GV/cm, vs cardinal number N (no units) in BaF.

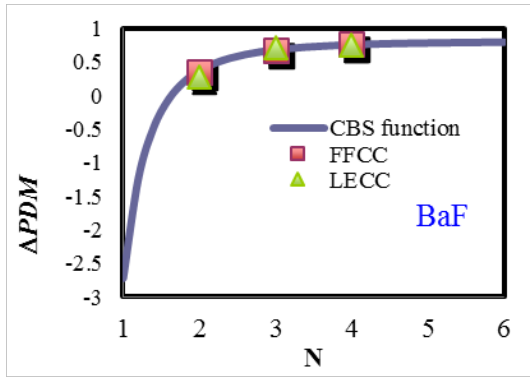

Figure 10. Correlation part of PDM ($\Delta$PDM), in Debye, vs cardinal number N (no units) in BaF.

We shall now briefly mention the previous works on the PDMs of the AEM systems (except BaF), in Table IV. We only provide the results and references, since a detailed comparison has been done in Ref. [13].

Table IV: Comparison of the PDMs of AEMs, with previous works. The expansions for the abbreviations in the Table are discussed in Ref. [13].

| Molecule | Work | Method | PDM |
|---|---|---|---|
| **BeF** | Langhoff et al [39] | CPF | 1.086 |
|  | Buckingham and Olegario [40] | MP2 | 1.197 |
|  | Kobus et al [41] | FD-HF | -1.273 |
| | Prasannaa et al [13] | LECCSD | 1.10 |
| | This work | FFCCSD | 1.15 |
| **MgF** | Torring et al [42] | Ionic model | 3.64 |
| | Langhoff et al [39] | CPF | 3.078 |
| | Mestdagh and Visticot [43] | EPM | 3.5 |
| | Buckingham and Olegario [40] | MP2 | 3.186 |
| | Kobus et al [41] | FD-HF | -3.101 |
| | Prasannaa et al [13] | LECCSD | 3.06 |
| | This work | FFCCSD | 3.13 |
| **CaF** | Torring et al [42] | Ionic model | 3.34 |
| | Rice et al [44] | LFA | 3.01 |
| | Langhoff et al [39] | CPF | 3.06 |
| | Mestdagh and Visticot [43] | EPM | 3.2 |
| | Bundgen et al [45] | MRCI | 3.01 |
| | Allouche et al [46] | LFA | 3.55 |
| | Buckingham and Olegario [40] | MP2 | 3.19 |
| | Kobus et al [41] | FD-HF | -2.645 |
| | Prasannaa et al [13] | LECCSD | 3.16 |
| | This work | FFCCSD | 3.19 |
| | Experiment [47] | | 3.07(7) |
| **SrF** | Torring et al [42] | Ionic model | 3.67 |
| Langhoff et al [39] | CPF | 3.199 | |
| Mestdagh and Visticot [43] | EPM | 3.59 | |
| Allouche et al [46] | LFA | 3.79 | |
| Kobus et al [41] | FD-HF | -2.576 | |
| Sasmal et al [48] | Z-vector | 3.45 | |
| Prasannaa et al [13] | LECCSD | 3.6 | |
| This work | FFCCSD | 3.62 | |
| Experiment [49] | | 3.4676(1) | |

Table V provides a list of previous works on the eEDM candidates, BaF, HgX and PbF. We do not discuss the previous works on the PDM of BaF, since we have done it in Ref. [13]. We do not elaborate on HgX either, since it has been done in detail, in Ref.s [12] and [22], for the effective electric fields and PDMs respectively.

Table V: Comparison on the effective electric fields and PDMs of HgX and PbF, with previous works. The abbreviations used in the Table are expanded: EPM: electrostatic polarization model, LFA: ligand field approach, CI: configuration interaction, SCF: self-consistent field, ECP: effective core potential, prefix RAS: restricted active space, prefix CAS: complete active space, and prefix MR: multireference.

| | Work | Method | PDM (D) | $E_{eff}$ (GV/cm) |
|---|---|---|---|---|
| **BaF** | Torring et al [42] | Ionic Model | 3.44 | |
| | Mestdagh and Visticot [43] | EPM | 3.4 | |
| | Allouche et al [46] | LFA | 3.91 | |
| | Tohme and Korek [50] | MRCI | 2.96 | |
| | Prasannaa et al [13] | LECCSD | 3.4 | |
| | Experiment (PDM) [51] | | 3.17(3) | |
| | Kozlov et al [52] | ECP-RASSCF | | 7.5 |
| | Nayak and Chaudhuri [53] | RASCI | | 7.28 |
| | Meyer et al [54] | MRCI | | 5.1 |
| | Meyer et al [55] | MRCI | | 6.1 |
| | This work | FFCCSD | 3.41 | 6.46 |
| **HgF** | Yu Yu Dmitriev et al [56] | CI | 4.15 | 99.26 |
| | Prasannaa et al [22] | LECCSD | 2.61 | |
| | Meyer et al [54] | MRCI | | 68 |
| | Meyer et al [55] | MRCI | | 95 |
| | Prasannaa et al [12] | LECCSD | | 115.42 |
| | This work | FFCCSD | 2.92 | 116.37 |
| **HgCl** | Wadt [57] | CI | 3.28 | |
| | Prasannaa et al [22] | LECCSD | 2.72 | |
|     | Prasannaa et al [12]  | LECCSD     |      | 113.56      |
|     | This work             | FFCCSD     | 2.96 | 114.31      |
| **HgBr** | Wadt [57]        | CI         | 2.62 |             |
|     | Prasannaa et al [22]  | LECCSD     | 2.36 |             |
|     | Prasannaa et al [12]  | LECCSD     |      | 109.29      |
|     | This work             | FFCCSD     | 2.71 | 109.56      |
| **HgI** | Prasannaa et al [22] | LECCSD    | 1.64 |             |
|     | Prasannaa et al [12]  | LECCSD     |      | 109.3       |
|     | This work             | FFCCSD     | 2.06 | 109.56      |
| **PbF** | Kozlov et al [58]    | ECP-SCF    | 6.1  | 28.95       |
|     | Yu YuDmitriev et al [56] | ECP-SCF | 4.62 | 20.68,37.22 |
|     | Baklanov et al [59]   | ECP-CASSCF | 4.26 | 33.09       |
|     | Sasmal et al [60]     | Z-vector   |      | 38.2        |
|     | This work             | LECCSD     | 3.72 | 37.24       |
|     | This work             | FFCCSD     | 3.88 | 37.91       |
|     | Experiment [61]       |            | 3.5(3) |           |

We shall shortly mention some of the previous applications of the FFCC method to $E_{eff}$. The first application of the FFCC method to $E_{eff}$, to the best of our knowledge, was by Isaev et al [62]. They applied the technique to calculate $E_{eff}$ of $HI^+$. The method was also applied, to compute $E_{eff}$s of RaF and HfF+ systems, by Kudashov et al [63], and Skripnikov [64], respectively.

We now discuss elaborately the error estimates in our calculations. They mainly stem from three sources, (1) the choice of basis sets, (2) incompleteness of correlation effects in the wave function, and (3) neglecting the higher order terms in the CCSD expectation value. The errors from the last source were already discussed in the above sections in comparing the LECC and FFCC approaches. We shall now look at the errors from the first two sources. Table VI provides the error estimates for AEMs. The column, 'Source 1', refers to the percentage fraction difference between the finite field QZ and the TZ results. Assuming that the results do not change significantly beyond the QZ basis, this column provides the errors due to the choice of basis sets. The results for BaF would not be very reliable, since the results do not converge from DZ through QZ. We, therefore, do not present them here. Also, note that the errors in $E_{eff}$s of BeF and MgF are large, but this may be due to the TZ and QZ values being extremely small. The next column, 'Source 2', refers to the percentage fraction difference between FF-CCSD(T) and FF-CCSD results (using the same basis sets, at the QZ level). The finite field calculations are done for a sample perturbation parameter of $10^{-3}$. The FF-CCSD(T) results for BeF, MgF, CaF, and SrF PDM are (in Debye), respectively: 1.22, 3.19, 3.24, and 3.66, while for $E_{eff}$, they are (in GV/cm): 0.005, 0.07, 0.28, and 2.15 respectively. It is clear from the data that CCSD is indeed a very good approximation in these kinds of systems at QZ level of basis, and we need not go to CCSD(T) approximation, since the change in results is not significant enough to invest in large computational time.

Table VI: Estimated errors (%) in AEMs.

| Molecule | Property | Error (%) | |
|---|---|---|---|
| | | (1) Basis sets | (2) Correlation effects |
| BeF | $E_{\text{eff}}$ | 20.0 | 0.0 |
| | PDM | 2.6 | 5.7 |
| MgF | $E_{\text{eff}}$ | 14.3 | 0.0 |
| | PDM | 1.6 | 1.9 |
| CaF | $E_{\text{eff}}$ | 3.6 | 0.0 |
| | PDM | 0.6 | 1.5 |
| SrF | $E_{\text{eff}}$ | 1.8 | 0.5 |
| | PDM | 4.4 | 1.1 |

Table VII presents the error estimates for HgF. Since the HgX, and PbF, systems require large computational time, we only look at HgF as the simplest representative case for this analysis. In the Table, 'Source 1-a' refers to percentage fraction difference between the finite field TZ and the DZ results, while 'Source 1-b' refers to percentage fraction difference between HgF with diffuse functions, added to both the atoms, and our DZ calculations from Table II. This was not included for the AEMs, since we already add diffuse functions from Sapporo basis, for Sr and Ba. Adding diffuse functions to F for any AEM does not change the results significantly [65]. Adding them to the lighter alkaline earth metals does not affect $E_{\text{eff}}$. However, they contribute to around 3 percent to the PDM. Finally, 'Source 2' is the same as that in the previous Table, estimated from the difference between the FF-CCSD(T) and FF-CCSD methods at the DZ basis sets. The values obtained at the DZ FF-CCSD(T) level for $E_{\text{eff}}$ and PDM respectively, are: 120.31 GV/cm and 3.36 D, while for TZ, they are: 122.97 GV/am and 3.41 D. These results for HgF clearly indicates that while DZ and CCSD results still give a reasonable estimate for $E_{\text{eff}}$, we need to go beyond them for an accurate value of PDM.

We also have not given the total estimate, since the errors do not add up linearly. HgF is an example, where errors due to sources 1a, and 2, add linearly to about 6 percent for $E_{eff}$. However, the percentage fraction difference between FFCCSD(T) TZ and its DZ counterpart is only 2.1 percent. Similarly, for the PDM, it is only 2.05 percent.

Table VII: Estimated errors (%) in HgF.

| Molecule | Property | Error | | |
|---|---|---|---|---|
| | | (1-a) Basis sets | (1-b) Basis sets (diffuse) | (2) Correlation effect |
| HgF | $E_{eff}$ | 3.02 | -2.4 | 3.3 |
| | PDM | 2.67 | 1.2 | 13.1 |

In an earlier work, Sasmal et al had applied a derivative approach known as the coupled cluster Z vector method to obtain $E_{eff}$ and PDM of PbF [60], and also SrF [48], in the CCSD approximation. It is evident from Tables I, and II, that their results are in good agreement with our FFCCSD calculations.

**Conclusion:**

We have employed a finite field approach to the RCC method, to evaluate effective electric fields and PDMs of several molecules, many of which are of interest in probing fundamental physics. In doing so, we have gone beyond our previous LECC approximation. Unlike the LECC approximation, the FFCC method offers the advantage of evaluating properties without having to truncate the expression anywhere. We performed both two point and six point calculations for all the AEM molecules, for various values of the perturbation parameter λ, and found that a two point formula is sufficient for our purposes. We also observe that the properties do not change within three decimal places, with change in the perturbation parameter. From the comparison between the FFCC and LECC methods, $E_{eff}$ does not change significantly when the higher order terms are neglected in expectation calculations. However, the PDM can be sensitive to this effect in some molecules, if we use the DZ basis sets. The largest change is found as much as twenty percent in HgI molecule. Therefore, for molecules whose

PDMs are not known experimentally, we need to carefully account for the correlation effects. We also perform CBS extrapolation on the correlation parts of both the properties for the AEM molecules and confirm that the QZ results are close to the values of CBS. We also present a detailed and rigorous analysis of errors from basis set convergence and correlation effects for electronic wave function. Our work also includes the finite field computation of the properties, using the relativistic CCSD(T) calculations, considered the gold standard of electronic structure calculations. The FFCC approach is applicable to a large range of properties, and therefore with some modification, it is possible to calculate not just $E_{\text{eff}}$ and PDM, but also other molecular properties that are needed to aid in search of new physics beyond the Standard Model.

**Acknowledgements:**

A major part of the computational results reported in this work were performed on the high performance computing facilities of IIA, Bangalore, on the hydra cluster.